\documentclass[prd,aps,amsmath,twocolumn,superscriptaddress,amssymb,preprintnumbers,reprint,nofootinbib]{revtex4-2}
\pdfoutput=1
\usepackage{graphicx,array}
\usepackage{hyperref}
\usepackage{color}
\usepackage{amsmath,amssymb,slashed,latexsym}
\usepackage{bm}
\usepackage{braket}
\usepackage{placeins}
\usepackage{adjustbox}
\usepackage{subcaption}
\usepackage{enumitem}
\def\beq{\begin{equation}}
\def\eeq{\end{equation}}
\def\be{\begin{equation}}
\def\ee{\end{equation}}
\def\bea{\begin{eqnarray}}
\def\eea{\end{eqnarray}}

\usepackage{array}
\newcolumntype{P}[1]{>{\centering\arraybackslash}p{#1}}
\newcolumntype{M}[1]{>{\centering\arraybackslash}m{#1}}
\usepackage{tabularray}
\UseTblrLibrary{booktabs}

\definecolor{darkblue}{cmyk}{1,0.4,0,0.3}
\definecolor{violet}{cmyk}{0,1,0,0.2}
\hypersetup{colorlinks, bookmarksnumbered, citecolor=darkblue, linkcolor=darkblue, pdfstartview=FitH, urlcolor=darkblue, linktocpage}

\renewcommand\k{{\bf k}}
\newcommand\q{{\bf q}}

\begin{document}

\preprint{MIT-CTP/5895}
\preprint{CERN-TH-2025-155}
\title{Effective Field Theory Constraints on Primordial Black Holes \\ from the High-Redshift Lyman-$\alpha$ Forest}

\author{\large Mikhail M. Ivanov}
\email{ivanov99@mit.edu}
\affiliation{Center for Theoretical Physics -- a Leinweber Institute, Massachusetts Institute of Technology, 
Cambridge, MA 02139, USA}
\affiliation{The NSF AI Institute for Artificial Intelligence and Fundamental Interactions, Cambridge, MA 02139, USA}

\author{\large Sokratis Trifinopoulos}\email{sokratis.trifinopoulos@cern.ch}
\affiliation{Center for Theoretical Physics -- a Leinweber Institute, Massachusetts Institute of Technology, 
Cambridge, MA 02139, USA}
\affiliation{Theoretical Physics Department, CERN, 1211 Geneva 23, Switzerland}
\affiliation{Physik-Institut, Universit\"at Z\"urich, 8057 Z\"urich, Switzerland
}%

\begin{abstract}
We present updated constraints on the abundance of primordial black holes (PBHs) dark matter from the 
high-redshift 
Lyman-$\alpha$ forest
data from MIKE/HIRES experiments. 
Our analysis leverages an effective field theory (EFT) description of the 1D flux power spectrum, allowing us to 
analytically predict 
the Lyman-$\alpha$
fluctuations on quasi-linear
scales from first principles. 
Our EFT-based likelihood enables robust inference across redshifts $z = 4.2–5.4$ and down to 
scales of 100 kpc,
within previously unexplored regions of parameter space for this dataset. We derive new bounds on the PBH fraction with respect to the total dark matter $f_{\rm PBH}$, excluding populations with $f_{\rm PBH} \gtrsim 10^{-3}$ for masses $M_{\rm PBH} \sim 10^{4}$--$10^{16} M_\odot$. This offers the leading constraint for PBHs heavier than $10^{9} M_\odot$ and highlights the Lyman-$\alpha$ forest as a uniquely sensitive probe of new physics models that modify the structure formation history of our universe.
\end{abstract}

\maketitle

\section{Introduction}
\label{sec:intro}

The large-scale structure (LSS) of the Universe is rapidly becoming a leading observational pillar of modern cosmology. It captures the evolution of tiny primordial fluctuations—visible at the time of photon decoupling in the cosmic microwave background (CMB)—into the web of galaxies and matter that we observe today~\cite{Frenk:2012ph}. Measurements of the matter distribution inferred by LSS observations enable robust tests of the standard cosmological model, while constraining a range of nonstandard scenaria for the formation and growth of cosmic structure. 

LSS probes such as galactic observations~\cite{2010MNRAS.404...60R,Banik:2019smi,DES:2021wwk} and CMB measurements~\cite{Planck:2018nkj,Planck:2018vyg,CMB-S4:2016ple,Buckley:2025zgh}, may constrain the matter-power spectrum at the permill level, at larger spatial scales but they lose sensitivity at wavenumbers above $0.5\,\text{Mpc}^{-1}$. On the other hand, the Lyman-$\alpha$ forest~\cite{Chabanier:2019eai} and ultraviolet luminosity functions~\cite{Sabti:2021unj}, while not achieving the same level of precision, can extend cosmological constraints to higher redshifts and smaller scales with wavenumbers up to $k \sim 10\,\text{Mpc}^{-1}$. 
A striking feature of these probes is their sensitivity to the small scale matter power spectrum which may contain
contributions 
from beyond the standard matter components. 
Such perturbations arise in certain new physics models including axion-like particles~\cite{Kobayashi:2017jcf,Irsic:2019iff,Feix:2019lpo,Feix:2020txt,Co:2021lkc,Hutsi:2022fzw,Gorghetto:2022ikz,Bird:2023pkr,Winch:2024mrt,Ellis:2025xju}, and primordial black holes (PBHs)~\cite{Afshordi:2003zb,Carr:2018rid,Inman:2019wvr,Murgia:2019duy,Liu:2022bvr,Gouttenoire:2023nzr,Khan:2025kag}. 

In this work, we focus on Lyman-$\alpha$ constraints on PBHs~\cite{Hawking:1971ei} with masses in the range $10^4$–$10^{10}\,M_\odot$. Even as a sub-percent component of dark matter, PBHs in this mass window can leave rich cosmological signatures~\cite{Carr:2020gox}. On the lighter end they could grow into the supermassive black holes found in galactic centers today via accretion and mergers~\cite{Bean:2002kx,Kawasaki:2012kn,Ali-Haimoud:2016mbv,Clesse:2016vqa,Clesse:2017bsw,Serpico:2020ehh}, or they may have formed already supermassive from the collapse of large primordial fluctuations~\cite{Nakama:2016kfq,Nakama:2017xvq}. In the latter case, they would be present at the onset of matter domination, inducing a scale-invariant ``isocurvature'' component in the power spectrum and acting as seeds for accelerated galaxy formation via the \emph{Poisson effect}~\cite{Meszaros:1974tb,Carr:1975qj,Afshordi:2003zb,Carr:2018rid,Inman:2019wvr}. Binaries formed from such PBHs could also produce distinctive gravitational-wave signatures in the $\rm nHz$ frequency band~\cite{Sasaki:2018dmp,Atal:2020yic,Depta:2023qst,Gouttenoire:2023nzr}.  For even larger masses, $M_{\rm PBH} \gtrsim 10^{12}\,M_\odot$, \emph{stupendously large PBHs} (SLABs) may exist as rare relics roaming the intergalactic medium~\cite{Carr:2020erq}.

The Lyman-$\alpha$ forest, observed through absorption features in quasar spectra, probes the distribution of neutral hydrogen in the redshift range $2 < z < 5$. The presence of primordial black holes can lead to a significant enhancement of the transmitted flux power relative to the standard $\Lambda$CDM prediction. The state-of-the-art constraints on the PBH parameter space from the Lyman-$\alpha$ forest are given in Ref.~\cite{Murgia:2019duy}, where a suite of hydrodynamic simulations was performed with the isocurvature perturbations incorporated into the initial conditions at $z=199$. The obtained $2\sigma$ upper bound is of the form $f_{\rm PBH} < (M_{\rm PBH} /60\,M_\odot)^{-1}$. However, their analysis applies for PBH populations with $f_{\rm PBH}\lesssim 0.05$, and because no simulations were performed for $f_{\rm PBH} M_{\rm PBH}>10^4$, the bound cannot be directly extrapolated to the superheavy regime $M_{\rm PBH} \gtrsim 10^5 M_{\odot}$~\cite{Carr:2020erq,Carr:2020gox}.

 Our approach takes advantage of the effective field theory (EFT) of the Lyman-$\alpha$ forest developed originally in~\cite{Ivanov:2023yla} (see also Refs.~\cite{Garny:2018byk,Garny:2020rom,Chen:2021rnb,Ivanov:2024jtl,deBelsunce:2024rvv,Chudaykin:2025gsh,He:2025jwp,deBelsunce:2025bqc,deBelsunce:2025edy}). This framework offers a perturbative and systematically controlled description of the flux power spectrum, expressed in terms of large-scale cosmological observables and local bias operators. 
 It accounts for nonlinear gravitational evolution and redshift-space distortions in a consistent bottom-up way~\cite{Desjacques:2016bnm,Ivanov:2019pdj,Ivanov:2022mrd}.
 Crucially, EFT enables 
 analytic predictions 
 in the regimes 
 where hydrodynamical simulations have not been applied, thereby probing regions of the PBH parameter space that have not been tested before.
As such, EFT provides an important complementary framework 
for modeling
the Lyman-$\alpha$ forest. 

\section{Supermassive Primordial Black Holes}
\label{sec:PBHs}

Even though on sufficiently large scales the PBH fluctuations are adiabatic, at smaller scales the fact that PBHs are discrete compact objects becomes important. Assuming no particular initial clustering at the epoch of PBH formation, their locations are random and thus their spatial distribution adheres to Poissonian statistics. As a result, a shot noise is generated in the total energy density which corresponds to isocurvature perturbations~\cite{Carr:2018rid}. The perturbations remain frozen during the radiation-dominated era~\cite{Carr:1974nx} but evolve linearly right afterwards (i.e. $z < z_{\rm eq} \approx 3400$).

For simplicity, we assume a monochromatic mass function, and the PBH populations are characterized by their mass, $M_{\rm PBH}$, and the fractional abundance $f_{\rm PBH}$. The matter power spectrum today is modified by the addition of a scale-independent component $P_{\rm iso}$~\cite{Afshordi:2003zb,Inman:2019wvr} 
\begin{align} \label{eq:power_spectrum}
    &P(k) = P_{\rm ad}(k) + P_{\rm iso}~, \notag \\
    & P_{\rm iso} \simeq  \begin{cases}
    (f_{\rm PBH} D_+(0))^2/\bar{n}_{\rm PBH}~,& \text{if } k\leq k_{\rm cut} \\
    0~,              & \text{otherwise}
\end{cases}~,
\end{align}
\noindent
where $P_{\rm ad}(k)$ is the adiabatic mode in $\Lambda$CDM, $\bar{n}_{\rm PBH} = f_{\rm PBH} \rho_{\rm crit,0} \Omega_{\rm DM}/M_{\rm PBH}$ is the co-moving average number density of PBHs, and $D_+(z)$ is the  isocurvature transfer function~\cite{Inman:2019wvr}.

Phenomenologically, the isocurvature term is truncated at the scale $k_{\rm cut}$ which is the scale where we expect the  Press-Schechter theory (PS)~\cite{1974ApJ...187..425P},
producing the constant PBH power above, to break down. 
In the absence of a reliable description of the transition between the linear and the non-linear regimes, the cut-off scale is chosen to be the inverse mean separation between PBHs, $\bar{k}_{\rm PBH} = (2 \pi^2 \bar{n}_{\rm PBH})^{1/3}$~\cite{Inman:2019wvr,DeLuca:2020jug,Hutsi:2022fzw}. Additionally, we stress that strong non-linear dynamics around the PBHs develop when the PBHs are very rare and they effectively evolve in isolation. This situation is known as the \emph{seed effect}~\cite{Carr:2018rid} and becomes dominant in the regime $f_{\rm PBH} \lesssim (1+z_{\rm coll})/z_{\rm eq}$, where $z_{\rm coll}$
is the redshift of the collapse
of the non-linear structure that 
embeds PBHs. 
Therefore, we restrict our discussion within the range of validity of the Poisson effect, which for the relevant redshifts corresponds to $f_{\rm PBH} \gtrsim  10^{-3}$.

\section{Analysis details}
\label{sec:analys}

\paragraph*{The dataset.} 
The Lyman-$\alpha$ forest probes the spatial fluctuations in the transmitted flux $F = e^{-\tau}$ of high-redshift quasar spectra, where $\tau$ is the optical depth due to absorption by neutral hydrogen in the intergalactic medium (IGM). The key observable is the one-dimensional (1D) flux power spectrum $P_F(k_\parallel, z)$, which quantifies correlations of flux fluctuations along the line-of-sight (LoS) direction in redshift space. This observable indirectly traces the matter distribution while encoding additional astrophysical effects such as redshift-space distortions (RSD).

For this study we focus on high-redshift Lyman-$\alpha$ data,
probing the linear matter power spectrum
on scales smaller than other experiments, 
which is advantageous for 
PBH constraints. 
We use the MIKE and HIRES spectrograph data~\cite{Viel:2013fqw,Esposito:2022plo}
in redshift bins $z=4.2, 4.6, 5.0, 5.4$ covering 
the wavenumbers $k/([\text{km/s}]^{-1})=0.01-0.08$,
corresponding to the 
comoving wavenumbers
$(0.64-11.2)\,h\,\text{Mpc}^{-1}$.
We combine overlapping 
MIKE/HIRES redshift bins in
a single data vector
using inverse covariance weights
as detailed in Supplemental Material.

\textit{EFT of the Lyman-$\alpha$ Forest.--}
To disentangle cosmological information from small-scale baryonic physics, we employ the effective field theory (EFT) of the Lyman-$\alpha$ forest~\cite{Ivanov:2023yla}. This framework constructs the most general perturbative expansion of the flux field in terms of long-wavelength operators consistent with the symmetries of the problem, and is valid up to the scale where the nonlinear effects 
become strong. 
Typically, this is where the density field becomes fully non-linear. In a $\Lambda$CDM
cosmology
at redshift $z\sim 5$, this  happens only at very small wavenumbers, $k_{\rm NL}\sim 100\,h\,\text{Mpc}^{-1}$,
by virtue of the approximate 
scale-invariance
of the primordial power
spectrum. In this setup,
the relevant non-linear scale
will be set by the Lyman-$\alpha$
forest gas scale $k_J\gtrsim 20\,h\,\text{Mpc}^{-1}$~\cite{Villasenor:2022aiy}.
This suggests that the EFT is applicable to the entire range 
of scales probed by MIKE/HIRES data.
(Inhomogeneities in the ionizing background, 
and patchy re-ionization can introduce 
new scales relevant for the EFT, but we will ignore this effect in this analysis.)

EFT builds on 
i) the $SO(2)$ rotational symmetry implies that the only allowed operators are scalars and gradients along the LoS (denoted by $\parallel$), with the most important component being the matter overdensity field $\delta_m$, and the gradient of peculiar velocity $\eta \equiv \partial_\parallel v_\parallel/(aH)$, respectively; ii) the equivalence principle dictates that the higher-order terms (and counterterms) can depend only on velocity gradients and tidal fields.
We define the flux contrast field as
\begin{align}
    \delta_F(\mathbf{x}) \equiv \frac{F(\mathbf{x})}{\langle F \rangle} - 1\,,
\end{align}
and in the rest frame of a neutral hydrogen cloud the expansion reads
\begin{align} \label{eq:delta_tracer}
    \delta_F(\mathbf{x}) =&~ b_1\,\delta_m(\mathbf{x}) + b_\eta\,\eta(\mathbf{x}) + \sum_i b_{\mathcal{O}_i}\,\mathcal{O}_i(\mathbf{x}) \\
    & + \delta_{\rm ctr}(\mathbf{x})+\epsilon(\mathbf{x})\,,
\end{align}
where $\mathcal{O}_i$ denote the higher-order composite operators, and $\epsilon$ encodes stochastic noise~\cite{deBelsunce:2025bqc}. The expression \eqref{eq:delta_tracer} is transferred in the observer’s frame using the RSD mapping~\cite{Desjacques:2016bnm}.
 The observable 1D flux power spectrum is obtained then by integrating the 3D flux power spectrum over hard momenta along the LoS,
\begin{align}
\label{eq:p1d}
    P^{\rm 1D}_F(k_\parallel, z) = \frac{1}{2\pi} \int_{k_\parallel}^\infty \mathrm{d}k\,k\,P_F^{\rm 3D}\left(k, k_\parallel, z\right)\,,
\end{align}
where $P_F^{\rm 3D}\left(k, k_\parallel, z\right)$ is given by Fourier transform of the two-point correlation function of the field $\delta_F$. At tree-level it can be expressed as a generalization of the Kaiser formula for galaxies~\cite{Kaiser:1987qv}
\begin{align} \label{eq:P_tree}
    P_{\rm tree}(k, z) = \left(b_1 - b_\eta \mu^2 \frac{d\ln D_+(z)}{d\ln a}\right)^2  P_{\rm}(k)\,,
\end{align}
where $\mu=k_\parallel/k$ and $D_+$ is the growth factor, while at 1-loop level we have additional contributions from the operators $\mathcal{O}_i$ in Eq. \eqref{eq:delta_tracer}
computed in Ref.~\cite{Ivanov:2024jtl}.
$\delta_{\rm ctr}\sim \nabla^2\delta_{\rm lin}$
captures higher 
derivative bias, 
baryonic effects,
Doppler broadening, 
and gas smoothing.
The 1D projection integral introduces additional sensitivity
to small scales which physically
corresponds to additional stochastic corrections of the form $P_{\rm 1D,~stoch}(k_\parallel) = \mathcal C_0+\mathcal C_1k_\parallel^2+\mathcal C_2k_\parallel^4+...$, which absorb corrections originating from small-scale effects. 

In total, our EFT model introduces 19 nuisance parameters for each redshift bin.
This corresponds to the most general line-of-sight tracer
of the matter field.
To reduce the freedom in the fit, 
we impose a simulation-based
relationship between parameters
following~\cite{Ivanov:2024jtl,deBelsunce:2024rvv}.
We use precision
EFT parameter measurements
from the state of the art ACCEL$^2$
simulation~\cite{Chabanier:2024knr}
to extract relationships (priors)
on the non-linear EFT bias parameters $b_{\mathcal{O}_i}(b_1)$ as functions of the linear bias~\cite{deBelsunce:2024rvv}. 
In order to be more conservative and avoid
extrapolation
errors, 
we only use the 
$z=4,5$ ACCEL$^2$ EFT parameter
measurements 
which cover the range of redshifts relevant for 
the MIKE/HIRES data. 
Note that in this approach $b_1$
controls both the scale-independent and scale-independent bias $b_{\mathcal{O}_i}$ via the simulation-based priors.
While this modeling approach  
produces an excellent fit
to MIKE/HIRES data without any
additional calibrations, 
we also marginalize
over the 1D stochastic counterterms
in order to be more conservative. 
Physically, this is equivalent 
to marginalizing over astrophysical
uncertainties, e.g. thermal 
broadening. 
Specifically, 
we marginalize the stochastic parameters  
over a Gaussian distribution 
that is centered
at the ACCEL$^2$ prior $\mathcal C_n(b_1)$ and
has a width estimated from the small-scale sensitivity of the EFT 
fit to the simulation data,
as is standard in particle physics. 
This results in a highly flexible 
conservative model which has 4 independent free
parameters in each redshift bin.
Details on our theory model
are provided in Supplemental 
Material.

Finally, we include the 
effect of PBHs by
feeding the modified
linear matter power spectrum in Eq. \eqref{eq:power_spectrum}
to the tree-level and loop 
EFT calculations.

\paragraph*{Analysis.} 

We perform our parameter inference using \texttt{MontePython}~\cite{Brinckmann:2018cvx,Audren:2012wb} MCMC engine, interfaced with a modified version of 
the EFT code~\texttt{CLASS-PT}~\cite{Chudaykin:2020aoj}. We fix the six standard cosmological parameters \{$\omega_\mathrm{b}$, $\omega_\mathrm{DM}$, 100$\theta_\mathrm{s}$, $\tau_\mathrm{reio}$, $\ln(10^{10}A_\mathrm{s})$, $n_\mathrm{s}$\}
to their Planck CMB best-fit values. 

\section{Results}

\begin{figure}[h!]
\centering
  \includegraphics[width=1\linewidth]{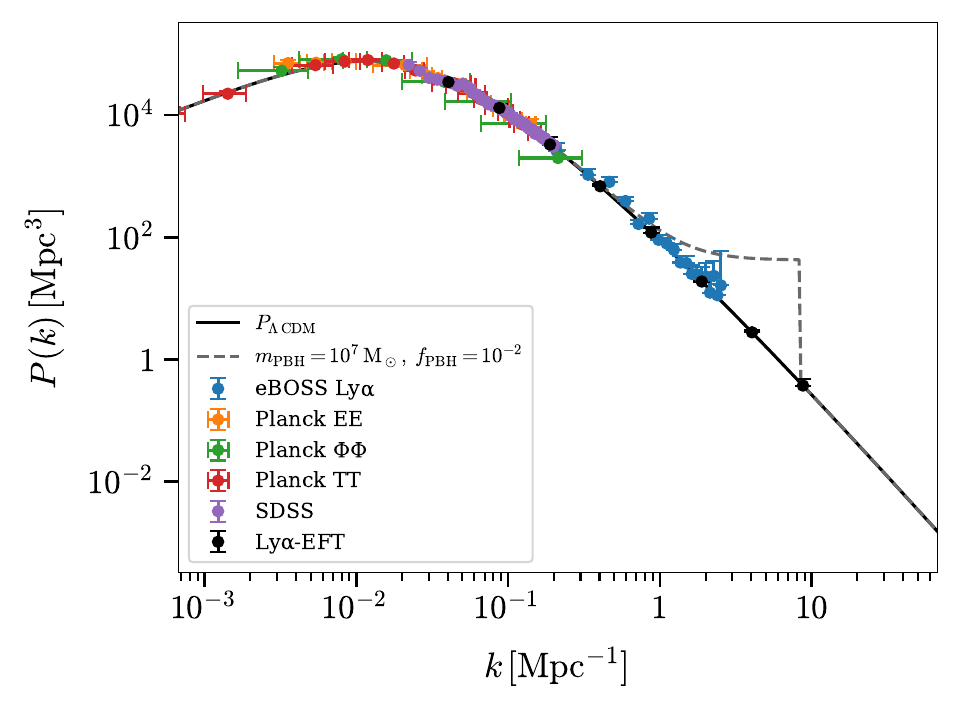} 
  \caption{The linear matter power spectrum at $z=0$ is shown for $\Lambda$CDM (black), and for a PBH model with $f_{\rm PBH}=10^{-2}$ and $M_{\rm PBH}=10^{7} M_{\odot}$ (gray dashed). The black dots (and corresponding errorbars) denote the 95\% C.L. constrains on overfluctuations from the high-redshift Lyman-$\alpha$ data from MIKE/HIRES, that are derived in this work. The blue dots denote the Lyman-$\alpha$ from eBOSS DR14 ~\cite{eBOSS:2018qyj,Chabanier:2019eai}. The rest of the errorbars correspond to observations of Luminous Red Galaxies (purple) by SDSS DR7 \cite{2010MNRAS.404...60R} (purple) and the {\it Planck} 2018 \cite{Planck:2018vyg} CMB temperature (red), polarization (orange) and lensing (green) power spectra.}
 \label{fig:Pkkcuts}
\end{figure}

Focusing on the new physics, we vary the constant term $P_{\rm iso}$ for different choices of $k_{\rm cut}/(h\text{Mpc}^{-1})$,
\begin{equation}
\begin{split} \label{eq:kcut}
   \{0.06,
0.13,
0.28,
0.60,
1.3,
2.8,
6.0,
13\} \,,
\end{split}
\end{equation} 
and later map the constraints on the PBH parameter space via eq.~\eqref{eq:power_spectrum}, e.g. the wavenumbers in Eq. \eqref{eq:kcut} correspond to the PBH masses $\log10(\{13,12,11,10,9,8,7,6\})\, M_{\odot}$
for $f_{\rm PBH} = 10^{-2}$.
The marginalized 95\% upper limits 
on $P_{\rm iso}/(h^{-1}\text{Mpc})^3$ for the corresponding $k_{\rm cut}$ read
\begin{equation}
\begin{split}
   \{1000,
660,
350,
20,
7.8,
1.6,
7.9\cdot 10^{-2},
3.4 \cdot 10^{-2}\} \,,
\end{split}
\end{equation} 
which can be readily 
converted into 
strong limits 
on the combination $f_{\rm PBH}M_{\rm PBH}$.
We stress that the results of our analysis can be readily applied to any model that predicts such a scale-invariant enhancement of the power-spectrum at the scales of interest. Our analysis constraints such departures from the $\Lambda$CDM matter power spectrum for $0.6 \lesssim  k/(h\, \rm{Mpc}^{-1})\lesssim 13\,$ at 
the $\lesssim 10\%$ level as can be seen in Fig.~\ref{fig:Pkkcuts}.

We stress that the
data points shown in Fig.~\ref{fig:Pkkcuts} are
correlated, which is particularly important
for $k<0.6~h$Mpc$^{-1}$ outside of the MIKE/HIRES range. The constraints on these wavenumbers come entirely from the EFT mode-coupling, which is 
an indirect effect, resulting in the abrupt deterioration of our constraints 
for $k_{\rm cut}<0.6~h$Mpc$^{-1}$. 
Note that the parameter space $k_{\rm cut}\lesssim 0.3~h$Mpc $^{-1}$ 
should be better
constrained by full-shape galaxy clustering
 data from DESI~\cite{DESI:2024hhd,Chudaykin:2025aux}.

\begin{figure*}[t!]
  \centering \includegraphics[width=0.8\linewidth]{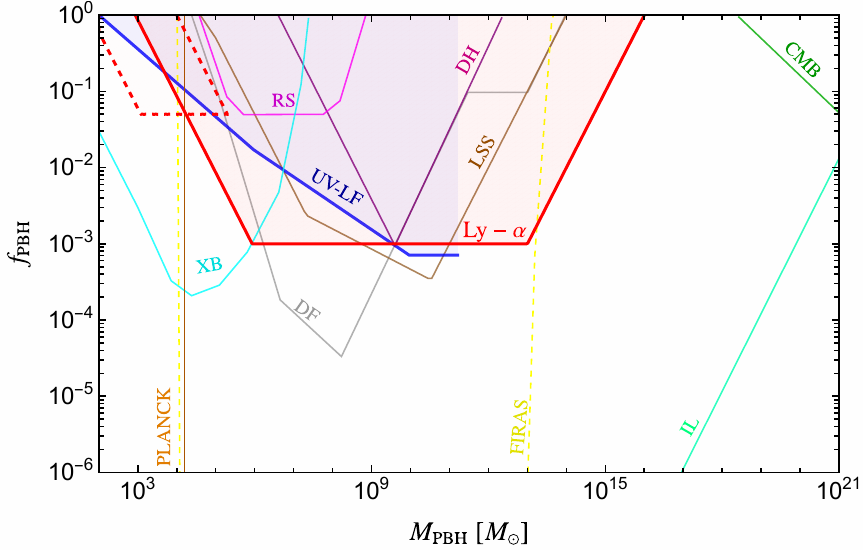} 
  \caption{Constraints on the parameter space of supermassive and stupendously large primordial black holes at 95\% C.L. The red shaded region denotes the region excluded due to Lyman-$\alpha$ MIKES/HIRES data derived in this work. We also show the Lyman-$\alpha$ constraints from the simulation-based study of Ref.~\cite{Murgia:2019duy} (dashed red). Limits arise also due to CMB anisotropies produced by accreting PBHs measured by Planck~\cite{Serpico:2020ehh} (PLANCK, yellow), X-ray flux induced by accretion~\cite{Inoue:2017csr} (XB, cyan), millilensing of radio sources~\cite{Wilkinson:2001vv} (magenta), dynamical friction causing halo objects to fall in the nucleus of the Milky Way~\cite{Carr:1997cn} (DF, grey), overheating of stars in the galactic disk~\cite{Carr:1997cn} (DH, purple), high-redshift formation of large-scale structures~\cite{Carr:2018rid} (LSS, brown),the Hubble Space Telescope Ultraviolet Luminosity Function~\cite{Gouttenoire:2023nzr,Sabti:2023xwo} (UVLF, blue), and the CMB dipole anisotropy~\cite{Carr:2020gox} (CMB, green). The region on the left side of the yellow dashed line is constrained by CMB spectral $\mu$-distortion measured by COBE/FIRAS surveys~\cite{Kohri:2014lza} in the Gaussian limit.\protect\footref{ftn:CMB_spectral} The right side of the light green line denotes the incredulity limit (IL), which corresponds to one black hole per Hubble volume.}
  \label{fig:PBH_space}
\end{figure*}

In Fig.~\ref{fig:PBH_space}, we show how those constraints translate on the 
$M_{\rm PBH}-f_{\rm PBH}$ plane together with all the other relevant constraints discussed in the literature. The constraint (red line) takes the shape of a trapezoid between $f_{\rm PBH}=1$ and the limit of the Poisson effect $f_{\rm PBH} \gtrsim \mathcal{O}(10^{-3})$ spanning a wide range of masses, i.e. between $10^{5}\,M_\odot$ and $10^{13}\,M_\odot$. The right side is determined by lower bin of $k_{\rm cut}$ in Eq. \eqref{eq:kcut}. 
Evidently, it becomes the leading constraint on the regime above $\mathcal{O}(10^{9}\,M_\odot)$\footnote{\label{ftn:CMB_spectral} Notice that the CMB spectral $\mu$-distortion bounds can be relaxed by introducing non-Gaussianities in the probability distribution of the primordial curvature perturbations~\cite{Nakama:2017xvq}.} stretching up to $10^{16}\,M_\odot$, deep into the region of SLABs. 
The left side corresponds to
the $P_{\rm iso}$ upper limit at $k_{\rm cut}=13~h$Mpc $^{-1}$ as all PBH populations with 
$k_{\rm cut}>13~h$Mpc $^{-1}$
have shot noise contributions
extending 
at least up to $k_{\rm cut}=13~h$Mpc $^{-1}$. 
In principle, we can get stronger 
constraints for 
\mbox{$k_{\rm cut}>13~h$Mpc $^{-1}$} as
EFT has sensitivity to 
these scales (outside of the MIKE/HIRES direct range) due 
to mode coupling, but we conservatively do not use this 
information here in order to be
more agnostic w.r.t. small-scale
dynamics.

We also show for comparison the previous state-of-the-art bound using the same dataset from Ref.~\cite{Murgia:2019duy}. A first observation is that the two bounds are probing complementary parts of the parameter space and feature different shapes. As expected, the simulation-driven analysis is capable of constraining the parameter space in the fully non-linear regime $k \gg k_{\rm NL}$, where the EFT description breaks down, and its shape is determined by the Poisson condition at the onset of the simulations $f_{\rm PBH} \gtrsim 0.05$. However, as explained in the Introduction, the reach of the bound for PBHs with masses above $10^5\,M_\odot$ is restricted by computational limitations. On the other hand, the EFT is applied at much lower redshifts, where the Poisson condition allows for smaller $f_{\rm PBH}$ values and successfully constraints scales that correspond to much heavier PBHs.

\section{Conclusions}

We have shown that the high-redshift Lyman-$\alpha$ forest data, interpreted through the EFT framework, can place stringent limits on new physics that enhances power at sub-halo scales by isocurvature fluctuations. 
EFT enables the efficient and computationally tractable analytic exploration of parameter spaces of such model.
As we argue here, a significant
fraction of the Lyman-$\alpha$ forest data at high redshift is approximately perturbative, and hence can be efficiently analyzed with EFT.
Focusing on the scenario of PBHs, we derived constraints on $f_{\rm PBH}$ in a regime previously unassessed from the state-of-the-art simulation-based studies. In particular, supermassive PBHs and SLABs with permill abundances with respect to dark matter, are decisively excluded. 

Ongoing and upcoming high-resolution spectroscopy of distant quasars and galaxies, e.g. the Lyman-$\alpha$ data from DESI~\cite{2013arXiv1308.0847L}
and Spec-S5 \cite{Spec-S5:2025uom}, 
are poised to grow the Lyman-$\alpha$ forest in the near future, providing 
more precision data to constrain small-scale
cosmological physics at high
redshifts. 
In addition, the galaxy clustering
data from DESI
analyzed with EFT
can be used to set limits 
on uncharted part of the 
parameter space of supermassive black holes.
These datasets together with other future galactic surveys, such as Euclid~\cite{Laureijs:2011gra}, the Rubin Observatory LSST~\cite{Abate:2012za}, as well as the 21-cm signal~\cite{Munoz:2019hjh} and even deep-field instruments like James-Webb Space Telescope (JWST)~\cite{2023Natur.616..266L,2024Natur.635..311X} will sharpen our understanding the of LSS structure in the early universe. These forthcomming observations offer thus a golden opportunity to go beyond not only the $\Lambda$CDM paradigm but also the Standard Model of particle physics.

\begin{acknowledgments}
We thank Mustafa Amin, M. Sten Delos, 
Vera Gluscevic, 
Naim G\" oksel Kara\c{c}ayl\i, Michael W. Toomey, and George Valogiannis for useful discussions.
We especially thank
Vid Ir\v{s}i\v{c}
for sharing the MIKE/HIRES data 
with us and for 
detailed comments
on the draft.
S.T. was supported by the Office of High Energy Physics of the U.S. Department of Energy (DOE) under Grant No.~DE-SC0012567, and by the DOE QuantISED program through the theory consortium “Intersections of QIS and Theoretical Particle Physics” at Fermilab (FNAL 20-17). S.T. is additionally supported by the Swiss National Science Foundation - project n. PZ00P2\_223581, and acknowledges CERN TH Department for hospitality while this research was being carried out.
\end{acknowledgments}

\bibliographystyle{JHEP}
\bibliography{biblio}

\newpage 

\pagebreak
\widetext
\begin{center}
\textbf{\large Supplemental Material}
\end{center}
\setcounter{equation}{0}
\setcounter{figure}{0}
\setcounter{table}{0}
\setcounter{page}{1}
\makeatletter
\renewcommand{\theequation}{S\arabic{equation}}
\renewcommand{\thefigure}{S\arabic{figure}}


\textit{Data combination.}
For $z=4.2,4.6,5.0$ where MIKE and HIRES data overlap,
we combine these datasets with the inverse covariance weights as
\be 
\bar P = C_{\rm comb} \cdot (C^{-1}_{\rm MIKE}P_{\rm MIKE} + C^{-1}_{\rm HIRES}P_{\rm HIRES})\,,\quad C_{\rm comb}\equiv (C^{-1}_{\rm MIKE} + C^{-1}_{\rm HIRES})^{-1}\,,
\ee 
where $P_{X}$ and $C_X$
are the data vectors 
and the covariance matrices 
for $X=$\{MIKE,HIRES\}.
For $z=5.4$ we only use the HIRES
datavector.

\textit{Modeling details.} 
The full EFT model for the 3D Lyman alpha correlations reads 
\be 
P^{\rm 3D}_F\left(k, k_\parallel, z\right)=(b_1-b_\eta f\mu^2)^2P(k) + P_{22} + P_{13} -2 k^2 P_{11}(c_0+c_2\mu^2+c_4\mu^4)\,,
\ee 
where $f=\frac{d\ln D_+}{d\ln a}$, $b_1(z),b_\eta(z)$
are linear bias parameters, $c_n$ $(n=0,2,4)$ are higher-derivative counterterms, and
\be 
\begin{split}
& P_{22}=\int 2\frac{d^3q}{(2\pi)^3}[K_2(\k-\q,\q)]^2P(|\k-\q|)P(q)~\,,\quad  P_{13} =6(b_1-b_\eta f\mu^2)P(k) \int \frac{d^3q}{(2\pi)^3}K_3(\k,-\q,\q)P(q)\,,
\end{split}
\ee 
and $K_n$
are the Ly-$\alpha$
EFT kernels~\cite{Ivanov:2023yla}.
We ignore the 3D stochastic contributions. They will be added with the 
appropriate counterterms
when we switch to 
1D correlations. Precision
measurements of EFT parameters from the ACCEL$^2$ simulation based on Ref.~\cite{deBelsunce:2024rvv} carried out
for the relevant redshift
bins $z=4,5$
imply the following relations between the EFT parameters:
\[
\begin{aligned}
b_\eta &= -0.46107\,b_1 + 0.22784,\quad 
b_{\eta^2}= -1.66391\,b_1 - 0.59993,\quad 
b_{\mathcal{G}_2}= -2.56067\,b_1 - 1.82742,\\
b_2 &= -10.01096\,b_1 - 3.23455,\quad
b_{\delta\eta}= -5.22520\,b_1 - 1.38922,\quad 
b_{(KK)_\parallel} = 6.13639\,b_1 + 2.43267,\\
b_{\Pi^{(2)}_\parallel} &= -0.99037\,b_1 + 0.17342,\quad
b_{\Pi^{(3)}_\parallel} = -0.35982\,b_1 + 1.32888,\quad
b_{\delta\Pi^{(2)}_\parallel} = -3.19383\,b_1 - 2.81066,\\
b_{(K\Pi^{(2)})_\parallel} &= -1.35666\,b_1 - 2.60276,\quad
b_{\eta\Pi^{(2)}_\parallel} = -10.61760\,b_1 - 6.93219,\quad
b_{\Gamma_3} = -0.11903\,b_1 - 0.24599\,,
\\
c_0 &= 0.02827\,b_1 + 0.01177,\quad
c_2 = -0.01933\,b_1 - 0.01002,\quad
c_4 = 0.00535\,b_1 + 0.00321\,.
\end{aligned}
\]
The uncertainties 
in these relations 
are negligibly small
thanks to an extended range of scales used in the fit at high redshifts, so we ignore them in what follows. 
We additionally multiply
our 3D EFT model with a 
Doppler broadening kernel,
\be 
P^{\rm 3D}_F\left(k, k_\parallel, z\right)\to P^{\rm 3D}_F\left(k, k_\parallel, z\right)e^{-k^2_\parallel/k_F^2}\,.
\ee 
We do not find 
any dependence 
on $k_F$ because this parameter is degenerate with out counterterm $c_2$, so 
we set $k_{F}=18
~h\text{Mpc}^{-1}
$
in what follows.
The 
1D projection integral 
in eq.~\eqref{eq:p1d}
is evaluated at some cutoff
$\Lambda\lesssim k_{\rm NL}$,
\begin{align}
    P^{\rm 1D}_F(k_\parallel, z) = \frac{1}{2\pi} \int_{k_\parallel}^\Lambda \mathrm{d}k\,k\,P^{\rm 3D}_F\left(k, k_\parallel, z\right)\,.
\end{align}
The sensitivity to 
small scales controlled by 
$\Lambda$ 
is renormalized by the 1D stochastic counterterms
\be 
P_{\rm 1D,~stoch}(k_\parallel) = \mathcal C_0+\mathcal C_1k_\parallel^2+
\mathcal C_2k_\parallel^4+...\,.
\ee 
In practice we use
$\Lambda=20~h$Mpc$^{-1}$
for the linear theory piece and 
$\Lambda=12~h$Mpc$^{-1}$
for the one-loop contribution. 
These choices are made
in order to be conservative
and ensure $\Lambda \ll k_{\rm NL}$, which is especially
important for the one-loop
correction which is more UV-sensitive
than the linear theory integrals. 
Within this scheme, the ACCEL$^2$ 1D data implies 
the priors 
\be 
\mathcal C_0 = 0.82540\,b_1 + 0.13270\,,\quad
\mathcal C_1= -0.00725\,b_1 - 0.00133\,,\quad
\mathcal C_2= 0.0000050\,b_1 + 0.0000004\,.
\ee 
In order to be conservative, 
we additionally marginalize these relations over 
some width
$\Delta  \mathcal C_n$,
estimated from the UV-sensitivity
of the theory predictions. 
We find that increasing the cutoff $\Lambda$
to $\Lambda'=30~h$Mpc$^{-1}$
(and switching off the Doppler broadening) 
shifts $\mathcal C_n$ by
\be 
\Delta  \mathcal C_n\simeq  
{0.05~h^{-1}\text{Mpc}}/{
[20~h\text{Mpc}^{-1}]^{2n}}~\,,
\ee 
which we choose as a prior on
$\Delta  \mathcal C_n$
that we marginalize over independently for every redshift bin. 

\begin{figure*}[t!]
  \centering \includegraphics[width=0.8\linewidth]{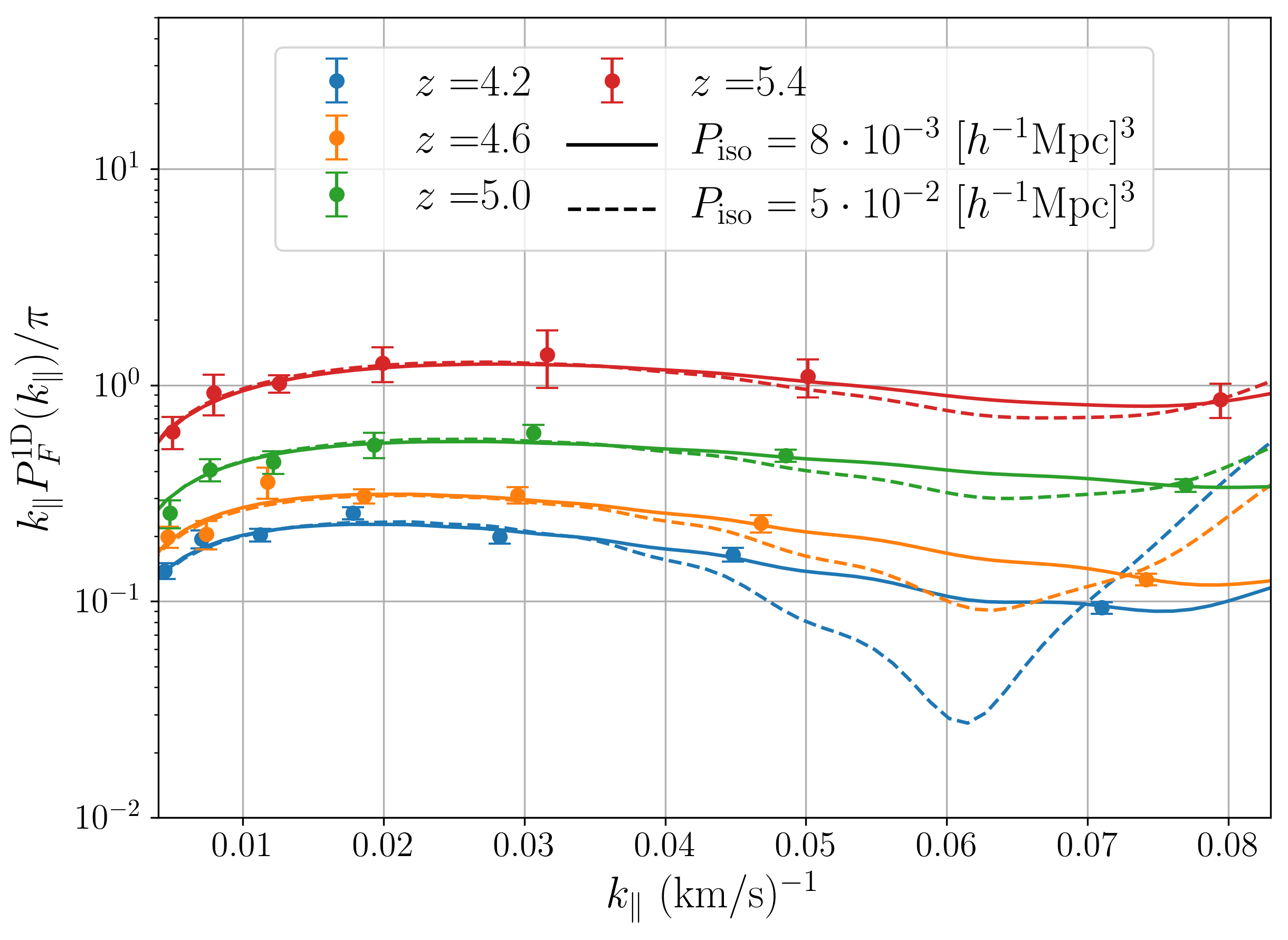} 
\caption{Combined MIKE/HIRES 1D flux power spectrum data and the best-fitting EFT theory curves. The data points are offset along the x axis for clarity. The error bars shown are diagonal elements 
of the covariance matrix. Note that they account for both 
statistical and systematic uncertainties. }
\label{fig:bf}
\end{figure*}

The best-fit to the 
combined MIKE/HIRES data 
in the PBH analysis 
with $k_{\rm cut}\simeq 13~h$Mpc$^{-1}$
and $P_{\rm iso}=8\cdot 10^{-3}$
[$h^{-1}$Mpc]$^3$
is shown in Fig.~\ref{fig:bf}.
Nominally, it has $\chi^2/\rm{d.o.f.}=1.09$ (recall that we have 4 fitting parameters per redshift bin consisting of 7 data points each), which indicates an excellent fit. 
Here it is important to note that the MIKE/HIRES covariance matrix includes
a multiplicative 
correction to account for possible systematic effects~\cite{Viel:2013fqw,Esposito:2022plo}.
In addition, we show the best-fit obtained for
fixed
$P_{\rm iso}=5\cdot 10^{-2}$
[$h^{-1}$Mpc]$^3$
and varied nuisance 
parameters. 
This value is ruled out by the data at about $3\sigma$, 
as can be clearly appreciated 
in the plot.
Note that the appearance of the 
dip in the PBH model around $0.06~[\text{km/s}]^{-1}$
is a non-linear effect due to the enhanced small-scale
displacements. 
This effect significantly reduces at high 
redshift, which is consistent with its non-linear nature. 
Curiously, this effect makes it harder to detect the PBH
signatures
by effectively
reducing 
its shot noise 
contribution.

\end{document}